# Introducing the Biomechanics-Function Relationship in Glaucoma: Improved Visual Field Loss Predictions from intraocular pressure-induced Neural Tissue Strains

Thanadet Chuangsuwanich[1,2], Monisha E. Nongpiur,[2,3] Fabian A. Braeu,[1,2,6] Tin A. Tun,[2,3] Alexandre Thiery,[4] Shamira Perera,[2,3] Ching Lin Ho,[2,3] Martin Buist,[5] George Barbastathis,[6,7] Tin Aung[1,2,3] and Michaël J.A. Girard[2,3,5,8,9,10]

[1]Yong Loo Lin School of Medicine, National University of Singapore, Singapore
[2]Singapore Eye Research Institute, Singapore National Eye Centre, Singapore.
[3]Duke-NUS Medical School, Singapore, Singapore.
[4]Department of Statistics & Data Science, National University of Singapore, Singapore
[5]Department of Biomedical Engineering, National University of Singapore, Singapore
[6]Singapore-MIT Alliance for Research and Technology, Singapore
[7]Department of Mechanical Engineering, Massachusetts Institute of Technology, Cambridge, Massachusetts, USA
[8]Department of Ophthalmology, Emory University School of Medicine, Atlanta, Georgia USA
[9]Department of Biomedical Engineering, Georgia Institute of Technology/Emory University, Atlanta, GA, USA
[10]Emory Empathetic AI for Health Institute, Emory University, Atlanta, GA, USA

**Corresponding Authors:**
Michaël J.A. Girard
Ophthalmic Engineering & Innovation Laboratory (OEIL), Singapore Eye Research Institute (SERI), The Academia, 20 College Road, Discovery Tower Level 6, Singapore.
E-mail: mgirard@ophthalmic.engineering

# Key Points

**Question:** How significant is the role of optic nerve head biomechanics in enhancing our understanding and analysis of the structure-function relationship in glaucoma?

**Findings:** We found that integrating biomechanical (intraocular pressure [IOP]-induced neural tissue strains) in addition to structural information significantly improves our prediction of the visual field loss.

**Meaning:** Our findings highlights the importance of the biomechanics-function relationship in glaucoma, and suggests that biomechanics may serve as a crucial indicator for the development and progression of glaucoma.

# Abstract

**Importance.** We demonstrated that incorporating biomechanics can enhance the structure-function relationship in glaucoma by increasing the accuracy of visual field loss predictions.

**Objective.** (1) To assess whether neural tissue structure and biomechanics could predict functional loss in glaucoma; (2) To evaluate the importance of biomechanics in making such predictions.

**Design, Setting and Participants.** We recruited 238 glaucoma subjects (Chinese ethnicity, more than 50 years old). For one eye of each subject, we imaged the optic nerve head (ONH) using spectral-domain OCT under the following conditions: **(1)** primary gaze and **(2)** primary gaze with acute IOP elevation (to approximately 35 mmHg) achieved through ophthalmo-dynamometry.

**Main Outcomes and Measures:** We utilized automatic segmentation of optic nerve head (ONH) tissues and digital volume correlation (DVC) analysis to compute intraocular pressure (IOP)-induced neural tissue strains. A robust geometric deep learning approach, known as Point-Net, was employed to predict the full Humphrey 24-2 pattern standard deviation (PSD) maps from ONH structural and biomechanical information. For each point in each PSD map, we predicted whether it exhibited no defect or a PSD value of less than 5%. Predictive performance was evaluated using 5-fold cross-validation and the F1-score. We compared the model's performance with and without the inclusion of IOP-induced strains to assess the impact of biomechanics on prediction accuracy.

**Results:** Integrating biomechanical (IOP-induced neural tissue strains) and structural (tissue morphology and neural tissues thickness) information yielded a significantly better predictive model (F1-score: 0.76 ± 0.02) across validation subjects, as opposed to relying only on structural information, which resulted in a significantly lower F1-score of 0.71 ± 0.02 ($p < 0.05$). Our subjects had a mean age of 69±5 years. Among them, 88 were female. The cohort included a wide range of glaucoma severity, with Mean Deviation (MD) values ranging from -1.8 (mild) to -25.2 (severe), and an average MD value of -7.25±5.05.

**Conclusion and relevance:** Our study has shown that the integration of biomechanical data can significantly improve the accuracy of visual field loss predictions. This highlights the importance of the biomechanics-function relationship in glaucoma, and suggests that biomechanics may serve as a crucial indicator for the development and progression of glaucoma.

# Introduction

In glaucoma, the loss of retinal ganglion cells and their axons leads to both structural and functional changes.[1, 2] Structural changes are typically assessed by measuring retinal nerve fiber layer (RNFL) thinning using optical coherence tomography (OCT), while functional changes are evaluated through visual field testing. To enhance our understanding of these assessments, the concept of 'structure-function relationship' was introduced, aiming to better predict the severity and progression of visual field defects based on structural data.[3, 4] However, this relationship has limitations, especially as early functional changes occur before substantial structural damage,[5] and in advanced stages where extensive RNFL thinning and the 'floor effect' can obscure the prediction of disease progression.[6]

We strongly believe that assessing structure alone does not adequately reflect the health of retinal ganglion cell axons, their vulnerability to injury, and the consequent risk of vision loss. Biomechanical forces, such as intraocular pressure (IOP),[7] cerebrospinal fluid pressure (CSFP),[8] and optic nerve traction,[9, 10] continuously distort neural tissues, creating deformations that occur every second of our lives. We hypothesize that these deformations could be indicative of axonal health. Our recent patient studies[10-15] have shown an association between IOP-induced tissue deformations in the optic nerve head (ONH) and visual field loss, underscoring a potential role for biomechanics in glaucoma, as corroborated and suggested by others.[16-21] However, the clinical utility of biomechanics in predicting visual field patterns and their progression has yet to be evaluated.

In this study, we wanted to introduce the concept of 'biomechanics-function relationship'. We aimed to assess whether neural tissue structure and biomechanics could predict functional loss in glaucoma; and to evaluate the importance of biomechanics in making such predictions. Our work was performed using geometric deep learning – an AI methodology designed to interpret complex information from 3D shapes.

# Methods

## *Subjects Recruitment*

We recruited 238 glaucoma subjects from clinics at the Singapore National Eye Centre. We included subjects aged more than 50 years old, of Chinese ethnicity (predominant in Singapore), with a refractive error of ±3 diopters, and excluded subjects who underwent prior intraocular/orbital/brain surgeries, subjects with past history of strabismus, ocular trauma, ocular motor palsies, orbital/brain tumors; with clinically abnormal saccadic or pursuit eye movements; subjects with known carotid or peripheral vascular disease; or with any other abnormal ophthalmic and neurological conditions; subjects with visual field deficits related to diabetic retinopathy or any other optic neuropathies or with advanced glaucoma (mean deviation of < –20 decibels [dB]) were excluded. Glaucoma was defined as glaucomatous optic neuropathy, characterized as loss of neuroretinal rim with vertical cup-to-disc ratio >0.7 or focal notching with nerve fiber layer defect attributable to glaucoma, with repeatable glaucomatous visual field defects (independent of the IOP value) in at least one eye.

This study was approved by the SingHealth Centralized Institutional Review Board and adhered to the tenets of the Declaration of Helsinki. Written informed consent was obtained from each subject.

## *Visual Field Testing*

Unreliable visual field test results with a false-positive error of more than 15% or a fixation loss of more than 33% were excluded. The thresholding algorithm (Swedish interactive testing algorithm) standard 24-2 program was selected. For every selected patient, the pattern standard deviation map from the Humphrey visual field report was used in the study. For each point in the pattern standard deviation (PSD) map, we defined a defect as a point having a p-value of less than 5%.

### *OCT Imaging and Biomechanical Testing*

We selected one eye at random from each patient, and we imaged the ONH with spectral-domain OCT (Spectralis; Heidelberg Engineering GmbH). The imaging protocol was similar to that from our previous work.[12] In brief, we conducted a raster scan of the ONH (covering a rectangular region of 150 × 100 centered at the ONH) comprising 97 serial B-scans, with each B-scan comprising 384 A-scans. The average distance between B-scans was 35.1 µm, and the axial and lateral resolution on average were 3.87 µm and 11.5 µm, respectively. Each eye was scanned twice under 2 conditions: OCT at baseline and under acute IOP elevation. Each patient was administered 1.0% tropicamide to dilate the pupils before imaging.

For each eye in baseline position, we applied a constant force of 0.65 N to the temporal side of the lower eyelid using an ophthalmodynamometer (ODM), as per a well-established protocol.[10, 12, 22] This force raised IOP to approximately 35 mmHg and was maintained constant. IOP then was reassessed with a TonoPen (Reichert Instruments GmbH), and the ONH was imaged with OCT immediately (within 30 seconds) after the IOP was measured.

### *AI-based Segmentation of ONH Tissues and Representation of the ONH Structure as 3D Point Cloud*

We automatically segmented all baseline OCT volume scans of the ONH using REFLECTIVITY (Reflectivity, Abyss Processing Pte Ltd, Singapore). More specifically, the following ONH tissue groups were automatically labelled (**Figure 1a**): (1) the RNFL and the prelamina tissue; (2) the ganglion cell layer and the inner plexiform layer; (3) all other retinal layers; (4) the retinal pigment epithelium; and (5) the OCT-visible part of the LC.

To prepare input for geometric deep learning, we represent the ONH structure (neural tissues + LC) as point clouds derived from the segmentation of OCT images. This process follows methods similar to those used in our previous work.[23-25] In brief, we first identified the anterior boundaries of all tissue layers in the segmented OCT scan. Each anterior boundary voxel was then represented as a 3D point cloud. Additionally, for each point, we extracted the local tissue thickness (minimum distance between anterior and posterior boundary – **Figure 1b)**. Each ONH point cloud was represented by approximately 20,000 points. For training the model, 3,000 points were resampled from each cloud (details provided below). All ONH point clouds were aligned with respect to Bruch's membrane opening (BMO) to ensure consistency. Briefly, we calculated the BMO center based on the extracted BMO points and fitted a plane to these points using a least squares approach. Subsequently, each point cloud was centered at the BMO center and rotated so that the normal vector of the BMO plane aligned with the axial direction of the scan. Finally, we performed a cylindrical crop (with a radius of 1.75 mm aligned with the axial direction) to further homogenize the data (**Figure 1b**).

***Deformation Tracking and IOP-induced Effective Strain Derivation***

To extract the 3D deformation of the ONHs under an acute elevation in IOP, we used a commercial digital volume correlation (DVC) module (Amira (version: 2020.3), Thermo Fisher Scientific, USA). Details about the preprocessing of the volumes and DVC algorithm can be found in our previous work.[12] From the deformation mapping, we extracted the strain of the ONH in terms of effective strain, which represents a local 3D deformation that takes into account both the compressive and tensile effects (the higher the compressive or tensile strain, the higher the effective strain). We then assigned to each point in the point cloud the value of effective strain (**Figure 1b**) using a K-nearest neighbor interpolation (with K = 5).

### *Prediction of Visual Field Maps Using Geometric Deep Leaning*

Geometric deep learning is an emerging field of AI that can exploit knowledge from complex 3D structures and shapes represented as 3D point clouds. These techniques (also known as Point-Net or dynamic graph convolutional neural network)[24, 25] are particularly well-suited for medical applications, as they enable us to effectively identify changes in organ structure and shape, providing valuable insights into the development and progression of pathologies. In our previous studies,[23-25] we employed Point-Net to model the optic nerve head (ONH) shape in 3D from OCT images. Our findings demonstrated that Point-Net, coupled with the point cloud representation of the ONH, can reliably diagnose glaucoma and identify critical tissue landmarks that contribute to the diagnosis.

In this study, we employed Point-Net to take the point clouds such as the one shown in **Figure 1b** as an input to predict the full visual field (52 points) defect map. We used the same architecture as in our previous publication,[24] with a modification to the output layer to output a tensor with a dimension of 52 by 1, ensuring a 1-to-1 correspondence between each element in the output tensor and each visual field point. For each point, the model outputs a probability value between 0 and 1, indicating the probability of a defect. We used a probability threshold of 0.5 to determine whether each point in the visual field belongs to the defect or non-defect class.

The network was trained using binary cross-entropy loss. We split the dataset into training (80%), validation (10%) and test (10%) sets. Models hyperparameters were selected based on the best binary cross entropy loss in the validation sets, and model performance was evaluated on the test sets. To assess the performance of the model, we reported the mean F1-score of the test sets across the 5-fold cross-validation process. The F1-score is commonly used in classification tasks to provide a single measure of a model's performance by combining precision and recall into a harmonic mean. It is particularly useful for imbalanced datasets, where accuracy can be misleading. For our task, where most visual field points belong to the non-defect class, the F1-score is a more suitable metric than accuracy, as it accounts for both the detection of defects (recall) and the accuracy of defect predictions (precision).

During the training process, we augmented the point clouds in each sample using various techniques: random cropping, random rotations, and random sampling of points (selecting a subset of 3,000 points from each point cloud). These parameters align with those in our previous work.[26] This augmentation enhances the model's generalizability to unseen datasets.

We independently trained two geometric deep learning models: **(1)** one that incorporated tissue thickness and effective strain information, and **(2)** another incorporating tissue thickness alone. We then compared the mean F1-score between the two scenarios using a t-test with a significance level of 0.05.

## Results

## Subjects' demographics

A total of 238 Chinese glaucoma subjects were recruited for the study, with a mean age of 69±5 years. Among them, 88 were female. The cohort included a wide range of glaucoma severity, with Mean Deviation (MD) values ranging from -1.8 (mild) to -25.2 (severe), and an average MD value of -7.25±5.05. A summary of the subjects' demographics is provided in **Table 1**.

## Incorporation of Effective Strains Improves Visual Field Prediction

Our models exhibited good performance, achieving F1-score scores of up to 0.81, with the capability to accurately depict the general pattern of visual field defects (**Figure 2a**). Furthermore, the integration of biomechanical information, specifically IOP-induced effective strains, resulted in a more robust model (F1-score score: 0.76 ± 0.02), in contrast to the model relying solely on structural (tissue thickness and tissue morphology) information (F1-score: 0.71 ± 0.02, $p < 0.05$, **Figure 2b**).

## Discussion

This work serves as an exploratory study intended to provide a proof of concept for the significance of ONH biomechanics in glaucoma by directly correlating biomechanical measurements with functional loss. Specifically, we aimed to develop a robust visual field prediction model leveraging the morphological and biomechanical characteristics of the optic nerve head (ONH). Although previous research has explored visual field prediction using ONH images,[27] tissue parameters,[28] or a combination of both,[29, 30] our study is pioneering in evaluating the utility of biomechanical information for functional prediction through geometric deep learning.

In this study, we found that incorporating biomechanical strain information significantly enhanced visual field loss prediction in glaucoma. The biomechanical theory of glaucoma suggests that IOP-induced strain may: (1) cause mechanical damage to retinal ganglion cells (RGCs),[31, 32] (2) disrupt microcapillary blood flow in the lamina cribrosa (LC), choroid, and retina,[33, 34] and (3) impede axoplasmic flow.[35] These IOP-induced mechanical stresses—stretch, compression, and shear—may translate into complex insults at the axon level, affecting axonal health and increasing the likelihood of damage and vision loss. Our previous study strengthened the biomechanics-function relationship by revealing a negative correlation between IOP-induced ONH effective strain and retinal sensitivity in glaucoma patients.[12] In this study, we advanced this understanding by utilizing a geometric deep learning approach, which captures local information on both biomechanical strains and tissue thickness using point cloud data, and demonstrated a direct correlation of the IOP-induced strains to the visual field loss. This method enables comprehensive visual field map prediction without relying on prior knowledge of the spatial correlation between structure and function, traditionally done using mapping between anatomical regions of the ONH and the corresponding local region of visual field loss (Garway-Heath map).[3]

Additionally, we achieved good prediction results with a relatively small sample size of 238 subjects, compared to other recent studies that required several thousand subjects.[28, 29] This underscores the robustness of the geometric deep-learning approach. Our models accurately predicted the main archetypes of visual field loss, such as superior arcuate loss and inferior loss (**Figure 1a**). However, the accuracy of our models decreased in early glaucoma cases (the last row of **Figure 1a**), where only a few points of visual field defect were present. In such cases, our models predicted the visual field map as having no defect. This issue may arise due to the large spread in mean deviation (MD) values (**Table 1**) relative to our sample size, resulting in relatively few training samples across the glaucoma spectrum (early, moderate, and severe). We believe that our model could be further improved with additional subjects, particularly those with early glaucoma.

Our study has several limitations. First, we classified each point on the visual field map as either a defect or non-defect, rather than regressing to PSD or mean deviation (MD) values. Despite this, we believe that binarizing each visual field point is appropriate given our relatively small sample size. Our results still demonstrate the model's capability to accurately identify visual field defect patterns and that incorporating biomechanical information could improve the prediction performance. Second, our overall method depends on the segmentation algorithm and the types of tissues that were segmented. AI-based segmentation can be prone to error especially in the presence of noisy scans.[36] Third, our method of acute IOP elevation via ophthalmo-dynamometry could introduce various uncertainties, such as slight variability in the location of force application or differences in the biomechanical properties of each eye. These issues have been extensively discussed and quantified in our previous publications.[10, 12] Fourth, we did not account for optical distortions that could arise from the OCT scans. Differences in distortion characteristics between the baseline scan and the IOP-elevated scans may occur due to variations in camera positioning relative to the eye. This remains an unsolved problem intrinsic to OCT imaging.

In conclusion, our study provides evidence for the link between biomechanics and functional loss in glaucoma through a geometric deep learning approach. We have demonstrated that integrating biomechanical data significantly improved the accuracy of visual field loss predictions. This study lays the groundwork for further investigations into the biomechanics-function relationship, which could eventually serve as a crucial indicator of glaucoma development and progression.

## Acknowledgments

We acknowledge funding from (1) the donors of the National Glaucoma Research, a program of the BrightFocus Foundation, for support of this research (G2021010S [MJAG]); (2) the "Retinal Analytics through Machine learning aiding Physics (RAMP)" project that is supported by the National Research Foundation, Prime Minister's Office, Singapore under its IntraCreate Thematic Grant "Intersection Of Engineering And Health" - NRF2019-THE002-0006 awarded to the Singapore MIT Alliance for Research and Technology (SMART) Centre [MJAG/AT/GB], (3) the National Medical Research Council, Singapore (MOH-000435) [TA] and (4) the NMRC-LCG grant 'TAckling & Reducing Glaucoma Blindness with Emerging Technologies (TARGET)', award ID: MOH-OFLCG21jun-0003 [MJAG].

| Characteristic | Values (mean ± standard deviation) |
|---|---|
| Age (year) | 69 ± 5 |
| Sex, female (%) | 37% |
| Axial Length (mm) | 24.3 ± 1.0 |
| Visual field, MD (dB) | -7.25 ± 5.05 |
| Pattern standard deviation (dB) | 7.19 ± 3.79 |
| Baseline IOP (mmHg) | 17.3 ± 2.9 |
| IOP (mmHg) with indentation^ | 34.5 ± 7.0 |

**Table 1**: Subjects' demographics.

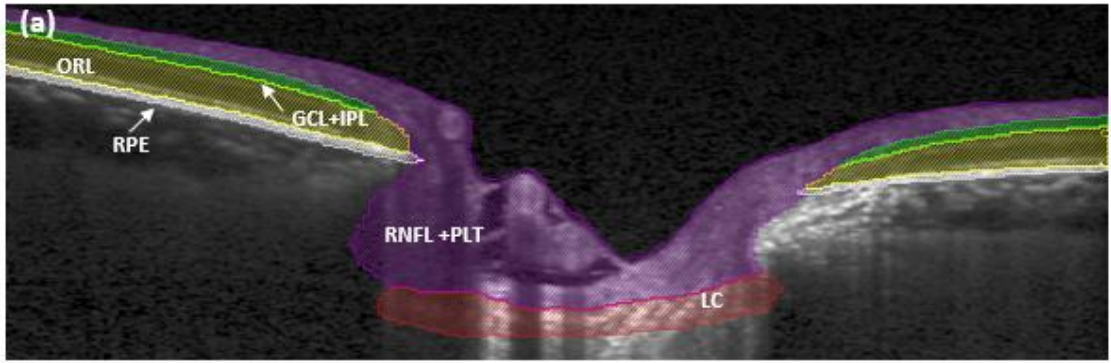

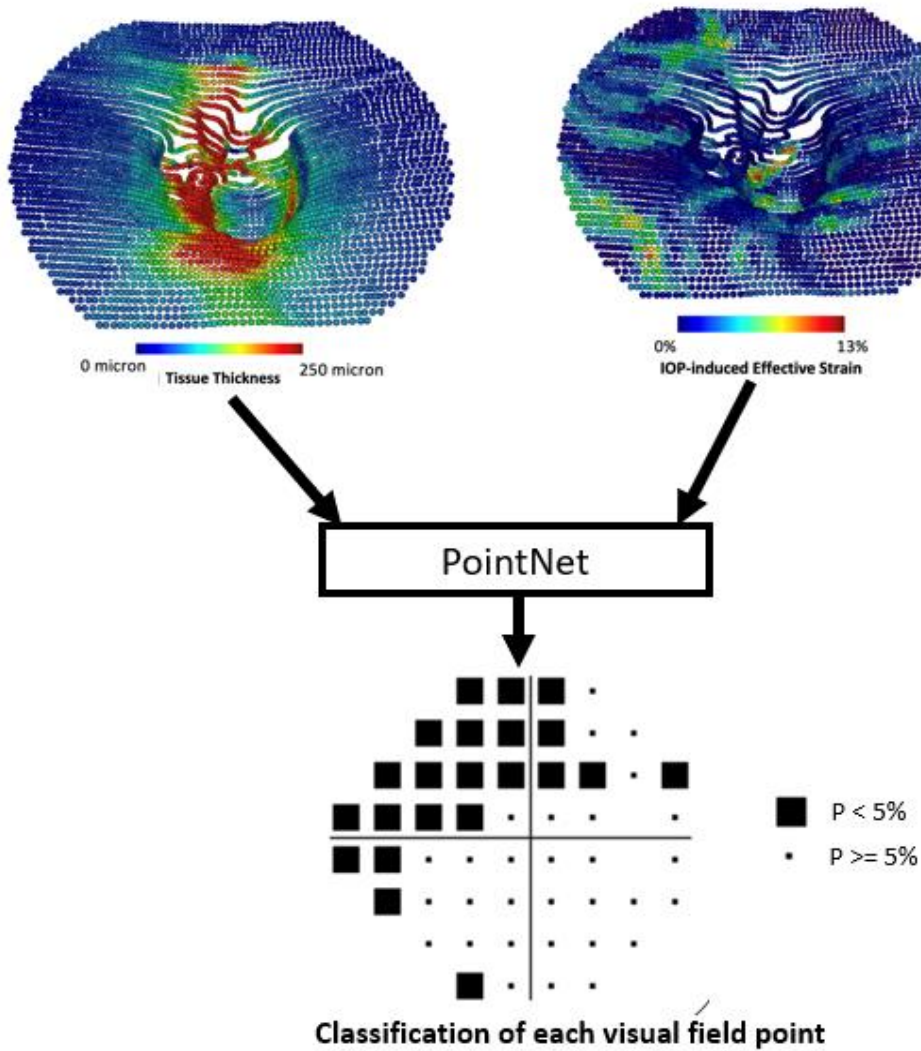

**Figure 1(a)** Segmentation of ONH neural tissues consisting of the retinal nerve fiber layer (RNFL) and the prelamina tissue (PLT), the ganglion cell layer and the inner plexiform layer (GCL+IPL), all other retinal layers (ORL), the retinal pigment epithelium (RPE) and the visible part of the lamina cribrosa (LC). **(b)** An illustration of ONH point

cloud containing both the tissue thickness and the effective strain as an input to the PointNet. The output is the 52 points visual field defect map.

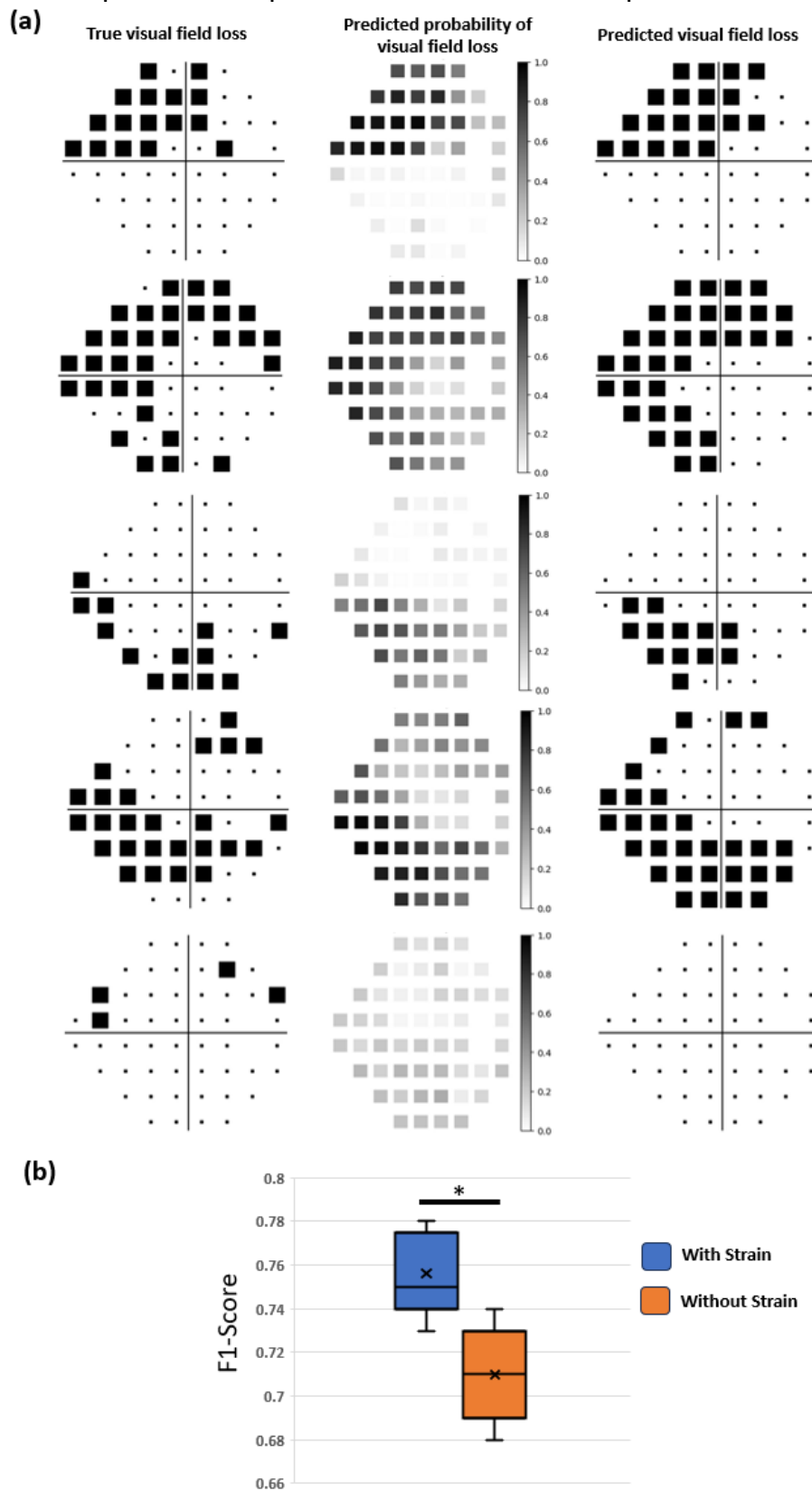

**Figure 2(a)** An example of a visual field defect prediction with both the probability values and final prediction after thresholding at 0.5. **(b)** A box plot showing a significant difference between the model with the inclusion of effective strain and the one without.